%Paper: hep-th/9204067
%From: aoki@madonna.physics.ucla.edu (Ken-ichiro Aoki)
%Date: Tue, 21 Apr 92 18:07:16 PDT

\input harvmac
\def\nl{\hfil\break}
\def\eg{{\it e.g.}}
\def\ie{{\it i.e.}}
\def\nextline{\hfil\break}

\def\npb{{  Nucl. Phys. }}

\def\rmp{{  Rev. Mod. Phys. }}
\def\half{{1\over2}}
\def\c#1{{\cal{#1}}}
\overfullrule=0pt
\def\comm#1#2{\left[#1,#2\right]}
\def\al#1{\alpha_{#1}}
\def\csa#1{h_{#1}}
\def\root#1{x_{#1}}
\def\str#1#2{c_{#1,#2}}
\def\cov#1{ D_{#1}}

\def\Om#1{\Omega_{#1}}

\nopagenumbers\abstractfont\hsize=\hstitle\rightline{UCLA/92/TEP/12}
\vskip 1in\centerline{\titlefont
W--Gravity and Generalized Lax Equations }
\centerline{\titlefont
for (super) Toda Theory}
\abstractfont\vskip .5in\pageno=0
\centerline{Kenichiro Aoki and Eric D'Hoker\footnote{$^\dagger$}
{Work supported in part by the National Science Foundation grant
NSF--89--15286.\nl
email:{\tt~aoki@uclahep.bitnet \rm and \tt dhoker@uclahep.bitnet}}}
\bigskip\centerline{\it Department of Physics}
\centerline{\it University of California at Los Angeles}
\centerline{\it Los Angeles, CA  90024{\rm --}1547}

\vskip .3in
\baselineskip=11.5pt plus 2pt minus 1pt
\centerline{\bf Abstract}
We generalize the Lax pair and B\"acklund transformations
for Toda and N=1 super Toda equations to the case of
arbitrary worldsheet background geometry.
We use the fact that the Toda equations
express constant curvature conditions, which arise
naturally from flatness conditions equivalent to
the W--gravity equations of motion.

\Date{4/92}
\newsec{Introduction}
The Toda chain problem in mechanics, and Toda field
theory have been known to be integrable systems
for some time now
\ref\rtoda{M. Toda, ``Studies of a non--linear Lattice"
Phys. Rep. 18C (1975) 1}%
\ref\leznov{A.N. Leznov and M.V. Saveliev, Lett. Math. Phys. 3 (1979) 207;
Comm. Math. Phys. 74 (1980) 111 \nextline
P. Mansfield, Nucl. Phys. B208 (1982) 277}%
\ref\Savel{M.V. Saveliev, Phys. Lett. A122 (1987) 312},
despite the fact that the precise
origin of this integrability has not heretofore
been elucidated.

In Liouville theory (the simplest Toda field theory),
integrability can be traced
to the fact that the field equation expresses
a constant curvature condition.
With constant curvature, the covering space of
any Riemann surface is A$_1$/U(1) and the constant curvature
geometry arises from the reduction of the flat
Maurer--Cartan form on the group A$_1$.\foot{By abuse of notation,
groups and algebras are not distinguished; \eg\ A$_1$ stands
for SU$(2)$ or any non--compact version of it.}
Thus the constant curvature condition is naturally equivalent
to a flatness condition on the A$_1$ connection, and
this is just the Lax pair
\ref\Liouv{E. D'Hoker, Phys. Lett. B264 (1991) 101}.
Thus, Liouville theory is intrinsically
related to the geometry of 2--dimensional Riemann surfaces, {\it i.e. }
gravity.
This geometrical framework for the understanding
of integrability allowed for the generalization
of the Lax and B\"acklund equations to
arbitrary background geometries
\Liouv
\ref\Thorn{C. Preitschopf and C. Thorn, Phys. Lett. B250 (1990) 79 }.

Within the context of 2-dimensional gravity alone,
there does not seem to be enough room for
an extension to include Toda field theory
as well.
We shall show in the present paper
that the correct geometrical framework for the
understanding of the integrability of Toda field theory is that of
2--dimensional W--gravity with gauge group a Lie group G.

In a topological field theory formulation of
W--gravity
\ref\Witten{
J.M.F. Labastida, M. Pernici and E. Witten,
Nucl. Phys. B310 (1988) 611\nl
D.~Montano, J. Sonnenschein, \npb 313 (1990) 258\nl
R.~Myers, V.~Periwal, \npb333 (1990) 536\nl
E. Witten, Nucl. Phys. B340 (1990) 281 \nextline
R. Dijkgraaf and E. Witten, Nucl. Phys. B342 (1990) 486}%
\ref\Montano{
D. Montano and J. Sonnenschein, Nucl. Phys. B234 (1989) 348}%
\ref\li{K. Li, Nucl. Phys. B346 (1990) 329 \nextline
K. Schoutens, A. Sevrin, P. van Niewenhuizen,
 Int. J. Mod. Phys. A6 (1991) 2891},
 the action is given by
\eqn\top{S=\int \tr({\cal N} {\cal F}) \qquad \qquad
{\cal F}= d{\cal A} + {\cal A}\wedge {\cal A}}
where ${\cal F}$ is the curvature of the
G-connection ${\cal A}$ and ${\cal N}$
is an auxiliary field.
Components of ${\cal A}$ corresponding to
generators of G with height $h$
have spin $|h|+1$.
Gauge transformations in G intertwine field
components of different spins, very much like
in supersymmetry theories.
We shall show that Toda field equations arise naturally as the
restriction of the W--gravity connection to the
case of components of height 1 only.
A flat connection always exists as
the Maurer--Cartan form on the group G,
and the Lax pair is the equation of
parallel transport on the surface.
In particular, we obtain Lax pair and
B\"acklund equations in the presence of
an arbitrary background geometry.

The above construction is generalized to the case
of W--supergravity as well.
For supergroups G where all simple roots can
be chosen to be of odd grading,
we derive the Toda Lax pair on arbitrary
supergravity background geometries.
For supergroups where some simple roots
must be even, supersymmetry is broken, and the
Toda field theory for these supergroups
is coupled only to background gravity, but
not supergravity.

Before we move on to describe Toda field theory
and its N=1 supersymmetric extension, we briefly recall some
basics on Lie (super) algebras.
We define a Lie algebra or a Lie superalgebra
(not necessarily finite dimensional)
by the following relations in the Chevalley basis
\ref\ALGEBRA{J.E.~Humphreys, {\sl``Introduction
to Lie algebras and representation theory"}, Springer--Verlag (1972);\nl
V.G.~Kac, {\sl ``Infinite dimensional Lie algebras"},
Cambridge University Press (1985);\nl
V.G.~Kac, {\sl Adv.~Math.} {\bf 26 } (1977) 8\nl
D.A. Leites, M.V.~Saveliev, V.V. Serganova, in {\sl ``Group
theoretical methods in physics",} VNU Science Press (1986)}.
We have
\foot{Throughout, $[.,.]$ denotes the graded commutator.}
\eqn\algebra{\left[\csa i,\csa j\right]=0,\quad
\comm{\csa i}{\root{\al j}}=k_{ij}\root{\al j},\quad
\comm{\csa i}{\root{-\al j}}=-k_{ij}\root{-\al j},\quad
\comm{\root{\al i}}{\root{-\al j}}=\delta_{ij}\csa i}
Here the generators of the Cartan subalgebra H
are denoted $h_i$ for $i=1,\dots,r={\rm rank}$ G,
the system of all roots is denoted $\Delta$
and $x_\alpha$ is the generator of G associated with the root
$\alpha \in \Delta$.
The system of (positive) simple roots is denoted $\Delta _s$,
and $k_{ij}$ is the Cartan matrix of G.
The full algebra is closed using the Jacobi identity
and the Serre relations, or equivalently
the restriction
\eqn\serre{\comm{\root\beta}{\root\gamma}
= \str\beta\gamma\root{\beta+\gamma} \qquad{\rm where}\quad\str\beta\gamma=0\
{\rm if}\ \beta+\gamma\not\in\Delta }
We may interpolate between the various compact and non--compact
versions of the algebra by rescaling the structure constants
$\str\beta\gamma$ using arbitrary real constants $\{\mu_i\}$ as
\eqn\contractions{\str\beta\gamma\mapsto\str\beta\gamma
\prod_i\mu_i^{\left(|\beta^i+\gamma^i|-|\beta^i|-|\gamma^i|\right)}}
where the components of $\gamma\in\Delta$ are defined by
$\gamma\equiv\gamma^i\al i$.
The height of a generator is defined as follows.
Simple positive roots have height 1, their negatives have height
$-1$ and the commutator preserves the grading.
For example, elements in the Cartan subalgebra H have height 0.
\newsec{Toda Theories for Ordinary Lie Groups Coupled to Gravity}
Two dimensional Riemannian geometry may be defined by the frame,
$e^a=d\xi^m e_m{}^a$,
and the U(1)--connection, $\omega = d\xi ^m \omega_m$.\foot{
Here, $\xi$ is a set of local
coordinates. We denote coordinate indices by $m,n,\ldots$
and the U$(1)$ frame  indices by
$a,b,\ldots$, where $a=z,\overline z$.
Also, $\delta_{z\overline z}=\delta_{\overline zz}=1$,
$\epsilon_{z\overline z}=-\epsilon_{\overline zz}=i$.  }
Covariant derivatives acting on tensors of weight $n$ are defined by
\eqn\der{D_a ^{(n)}= e_a{}^m (\partial _m + i n \omega _m )}
The metric is given by $g_{mn}=e_m{}^a e_n{}^b \delta _{ab}$.
As usual, the torsion and the curvature are defined
by the relation
\eqn\torsion{\comm{\cov a}{\cov b}=T_{ab}{}^c\cov c-in\epsilon_{ab}R_g}
Weyl transformations are defined by
\eqn\Weyl{e_m{}^a = \exp\{ \phi \} \hat e_m{}^a \qquad
\omega _m = \hat \omega _m + \epsilon _m{}^n \partial _p \phi}
under which torsion and curvature transform as
\eqn\Weyltrans{T_{ab}{}^c = \exp \{ -\phi \} \hat T _{ab}{}^c\qquad
R_g = R_{\hat g} \exp \{ -2\phi \} -2D_z D_{\bar z} \phi}
Upon setting $R_g$=constant in the last equation, we recover
the Liouville equation, for which a generalized Lax
pair was obtained in \Liouv\ in the following way.

Topological gravity is defined by an A$_1$ gauge field
\Montano
\eqn\sln{{\cal A} = -i \omega J_3 + e^z J_z + e^{\bar z} J_{\bar z}}
where the $J$'s are the generators of A$_1$.
Flatness of ${\cal A}$ is equivalent to zero torsion and constant curvature,
{\it i.e. } the Liouville equation, and the Maurer--Cartan form on A$_1$
always provides with such a connection.
The Lax pair is the equation for parallel transport.

Topological W--gravity is a construction similar
to that of topological gravity, but in which the
group A$_1$ is replaced with an arbitrary Lie group G.
Thus, we introduce a G--valued connection ${\cal A}$,
which may be decomposed as follows
\eqn\conn{\c A\equiv \sum _i
\omega^i \csa i+\sum_{\gamma\in\Delta}e^\gamma\root\gamma}
Here, $\omega ^i$ are the components of the Abelian connection
with gauge group H, and $e^\gamma$ are a generalization
of the frame on the Riemann surface $e^a$.
Actually, ${\cal A}$ may be viewed as a connection in the
bundle G with structure group H over the manifold G/H.
The latter is always a K\"ahler manifold, so the field
contents of W--gravity may be viewed as resulting from
embedding a Riemann surface into the K\"ahler manifold G/H
or equivalently from dimensional reduction of the
manifold G/H to a Riemann surface.
Analogous embedding problems and their relation to Toda systems
were considered in
\ref\gervais{J.-L. Gervais and Y. Matsuo, LPTENS--91/29, 35 (1991)}.

The corresponding gauge field strength is
\eqn\fieldtensor
{\eqalign{\c F      = & \sum _i d\omega^i\csa i +
\half \sum_{\gamma\in\Delta }e^\gamma \wedge e^{-\gamma}
\comm{\root\gamma}{\root{-\gamma}}\cr
& +\sum_{\gamma\in\Delta}\left[de^\gamma+\sum_{i,j}\omega ^i k_{ij}\gamma^j
\wedge e^\gamma +\half\sum_{\gamma'+\gamma''
=\gamma}e^{\gamma'}\wedge e^{\gamma''}\str{\gamma'}{\gamma''}\right]\root\gamma
\cr}}
The dynamics of topological W--gravity is given by action of \top\ and
reduces to $\c F =0$.
This produces ``torsion constraints" from the second line above and
constant curvature equations for the connections $\omega ^i$ from
the first line.
These are just the constant torsion and curvature
formulas for the H--connection on the manifold G/H.

Toda field theory is now simply obtained by restriction
to the spin 1 and spin 2 fields in the connection $\c A$ :
\eqn\ansatz{e^{\al i}=\exp\{\phi_i\}  e^z,\qquad
e^{-\al i}=\exp\{\phi_i\}  e^{\overline z},\qquad \qquad
e^\gamma=0\ \ {\rm if}\ |{\rm height (\gamma)}|\geq2}
as well as
\eqn\setconn{i\sum _j \omega ^j k_{ji}=\omega \rho _i +
e^a \epsilon_a{}^b  D_b\phi_i}
Using this expression, we see that the height of the field
is transformed into its spin.
Here $(e^a, \omega)$ describes an arbitrary 2--dimensional
Riemannian geometry as given above, and $\rho _i =1, \
i=1,\dots,r$ for this embedding.
Clearly, more general embeddings may be chosen, as can be seen
by applying an arbitrary W--gravity \ie\ G--gauge
transformations to Ansatz \ansatz.
The expression for the field strength reduces to
\eqn\fieldstrength{ \c F= \sum _ie^{\overline z}\wedge  e^z
\left[-(\cov z\omega_{\overline z}^i-\cov{\overline z}\omega_z^i
+\exp\{2\phi_i\}) \csa i
+\exp\{\phi_i\}  T_{z\overline z}{}^z\root{\al i}
+\exp\{\phi_i\}  T_{z\overline z}{}^{\overline z}\root{-\al i}\right]}
The W--gravity field equations $\c F=0$ are equivalent
to the zero torsion constraints of ordinary
gravity plus the curvature equations
\eqn\todaeq{2D_z D_{\bar z}\phi_i-\sum _j
\exp\{2\phi_j\} k_{ji}+R_{g}\rho _i=0}
which are precisely the Toda field equations.
When G is a Kac--Moody algebra, Cartan matrix $k$ has
an eigenvector $n_i$ with eigenvalue zero.
As a result, the linear combination $\sum n_i\phi_i$
obeys a linear equation. Upon elimination
of this field, conformal invariance is broken.
The Sine--Gordon system, for example, is obtained
in this way from G$=\hat {\rm A}_1$.

The Lax pair is identified as the equation for parallel
transport under the G--connection of G--valued functions $\psi$
in some representation of the group.
\eqn\laxpair{\left(\partial_m+\c A_m\right)\psi=0\quad\Leftrightarrow\quad
\left(\cov a+ \c A_a\right)\psi=0}
or in terms of frame index notation, using the explicit form of $\c A$
\eqn\lax{\eqalign{
(D_z + \sum_i\omega ^i _z h_i + \sum_i\exp \{ \phi _i \}
x_{\alpha _i} )\psi &=0 \cr
(D_{\bar z} + \sum_i\omega ^i _{\bar z} h_i +
\sum_i\exp \{ \phi _i \} x_{-\alpha _i} ) \psi &=0}}
These equations now provide a Lax pair for Toda field
theory on an arbitrary Riemann surface background geometry,
with the connections $\omega ^i$ given by \setconn.
Spectral parameters arise in the same way as they did
for Liouville theory.

When the Cartan matrix has an inverse $l_{ij}$ and is symmetric in
$i,j$, we may write an action
from which the Toda equations are obtained
\eqn\action{S_\phi=\int d^2\xi \sqrt g \sum _i \left [\sum _j \{ l_{ij}
D _z \phi _i D _{\bar z} \phi _j - R_g l_{ij} \rho _i \phi _j\}
+\half \exp\{ 2\phi _i\} \right ] }

{}From the Lax pair, we construct the B\"acklund transformation
by passing from homogeneous variables $\psi$ in the Lax equation
to projective (inhomogeneous) coordinates.
Saveliev \Savel\ has introduced a particularly natural way of
passing to inhomogeneous coordinates, including when
$\psi$ is in an arbitrary (finite-dimensional) representation
$\mu$ of G.

For a representation $\mu$ with highest weight vector $\mu$
all other weights can be built up by applying the generators
corresponding to negative
simple roots to the highest weight $|0;\mu\rangle $
\eqn\states{|j_1\dots j_p ;\mu\rangle
\equiv
x_{-\alpha _{j_p}} \dots x_{-\alpha _{j_1}} |0;\mu\rangle }
Here, it is always understood that $|j_1 \dots j_p;\mu\rangle  =0$
if the corresponding weight does not belong to the weight
diagram of $\mu$.
Application of Cartan generators and simple roots is
straightforward and may be found using the structure relations as
\eqn\basics{\eqalign{
h_j |j_1 \dots j_p ;\mu\rangle  =& \lambda ^{(p+1)} _{j;\mu}
|j_1 \dots j_p ;\mu\rangle  \cr
x_{-\alpha _j} |j_1 \dots j_p ;\mu\rangle  = &
|j_1 \dots j_p j ;\mu\rangle  \cr
x_{\alpha _j} |j_1 \dots j_p ;\mu\rangle  = &
\sum _{q=1} ^p \delta _{j,j_q} \lambda ^{(q)} _{j;\mu}
|j_1 \dots \hat j_q \dots j_p ;\mu\rangle  \cr}}
Here, the hat denotes omission and we shall use the abbreviation
\eqn\abb{\lambda ^{(q)} _{j;\mu}
\equiv \mu _j - \sum _{m=1} ^{q-1} k_{jj_m }}
We define the B\"acklund variables $\psi _{j_1 \dots j_p;\mu}$
through the matrix elements
\eqn\psis{\langle 0;\mu | \psi | 0 ;\mu\rangle
\exp \{ \psi _{j_1 \dots j_p;\mu} \}
=\langle j_1 \dots j_p ;\mu| \psi |0 ;\mu\rangle }
{}From the Lax pair \lax\ and equations \basics\ it is easy to see that
$\psi _{j_1 \dots j_p;\mu}$ satisfy
\eqn\Back{\eqalign{
D_z \psi _{j_1 \dots j_p;\mu} & + \sum _{i=1} ^r \left [
\lambda _{i;\mu} ^{(p+1)} \omega ^i _z + \exp \{
\phi _i + \psi _{j_1 \dots j_p i ;\mu}
- \psi _{j_1 \dots j_p;\mu}\} \right ] =0\cr
D_{\bar z} \psi _{j_1 \dots j_p;\mu} + & \sum _{i=1} ^r \left [
\lambda _{i;\mu} ^{(p+1)} \omega ^i _{\bar z} + \sum _{q=1} ^p
\delta _{i,j_q} \lambda _{i;\mu} ^{(q)} \exp \{
\phi _i + \psi _{j_1 \dots \hat j_q \dots j_p;\mu}-
\psi _{j_1 \dots j_p;\mu}\} \right ] =0 \cr }}
These are the B\"acklund transformations for Toda field
theory on an arbitrary background geometry and for
arbitrary (finite-dimensional) representations
of groups G.
For G=A$_n$, and $\mu$ the fundamental representation, we
get a B\"acklund transformation in the usual sense {\it i.e. }
there are as many $\psi$'s as $\phi$'s, and the integrability
on both are the Toda equations.
It is not known to us whether for other groups or other representations,
the above B\"acklund system can be further reduced; we suspect it
cannot in general.

\newsec{Toda theories coupled to supergravity}
Two dimensional supergeometry may be defined by the zweibein,
$E^A=d\xi ^ME_M{}^A$,
and the U(1)--connection, $\Omega = d\xi ^M\Omega_M$.
The covariant derivative acting on a superfield of U(1)
weight $n$ is given by
\eqn\covdef{\c D ^{(n)}_ A\equiv E_A{}^M\left(\partial_M+in\Om M\right)}
The torsion and the curvature tensors are defined by
\eqn\tordef{\comm{\c D _A}{\c D_ B}=T_{AB}{}^C\c D _ C+inR_{AB}}
acting on a superfield of weight $n$.
The standard torsion constraints are imposed
\ref\RMP{ P. Howe, J. Phys. A12 (1979) 393\nextline
J. Wess and J. Bagger, {\sl ``Supersymmetry and Supergravity",}
Princeton (1983) \nextline
E.~D'Hoker, D.H.~Phong, \rmp{\bf 60} (1988) 917}.

Topological W--supergravity is defined with respect to a supergroup G,
and an associated G--valued connection $\c A$.
This construction generalizes topological supergravity
which is constructed on the supergroup B(0,1)=OSp(1,1)
\ref\susy{D. Montano, K. Aoki and J. Sonnenschein,
Phys. Lett. B247 (1990) 64 }.
We may parametrize this connection as
\eqn\sconn{\c A\equiv \sum _i \Omega ^i \csa i+\sum_{\gamma\in\Delta}
E^\gamma\root\gamma}
The corresponding gauge field strength is
\eqn\sfieldtensor{\eqalign{
\c F = & \sum _i d\Omega ^i\csa i-\half \sum_{\gamma\in\Delta}E^{-\gamma}
\wedge E^\gamma \comm{\root\gamma}{\root{-\gamma}}\cr
& +\sum_{\gamma\in\Delta }\left[dE^\gamma+\sum_{i,j}\Omega ^i
k_{ij}\gamma^j
\wedge E^\gamma -\half\sum_{\gamma'+\gamma''=\gamma}
E^{\gamma'}\wedge E^{\gamma''}\str{\gamma''}{\gamma'}\right]\root\gamma}}
The equations of motion of topological supergravity are the
vanishing of this curvature $\c F =0$, and may again be viewed
as constant torsion and curvature equations of the
H--connection $\Omega ^i$ on the supermanifold G/H.

Two classes of supergroups must be distinguished.
First, we have supergroups for which {\it all} simple
positive roots may be chosen to have odd grading.
In the Kac classification, the
finite dimensional groups that have this property are
A$(n,n-1)$,\ B$(n-1,n)$,\ B$(n,n)$,\
D$(n+1,n)$,\ D$(n,n)$,\ D$(2,1;\alpha)$ \ALGEBRA.
For these groups, the corresponding Toda superfield
theory can be covariantly coupled to N=1 supergravity.
Second, there are all the other supergroups for which at
least one simple root must have even grading.
For these supergroups, Toda field theory is not
supersymmetric and can be coupled covariantly only to
ordinary gravity.
This distinction originates from Toda field theory on flat
superspace
\ref\STODA{M.A. Olshanetsky, Comm. Math. Phys. 88 (1983) 1205\nl
J. Evans and T. Hollowood, Nucl. Phys. B352 (1991) 723\nl
T. Inami and H. Kanno, Comm. Math. Phys. 136 (1991) 543}%
where global supersymmetry is preserved only in the first case
(see \STODA\ and references therein).

We treat the first case first, and use the convention
\eqn\cconvention{\str{\al i}{\al j}=\str{-\al i}{-\al j}=2,\quad
{\rm for\ }\forall\al i,\al j\in\Delta_s}
We recover N=1 Toda field theory by  making the following reduction
\eqn\sansatz{\eqalign{
E^{\al i}&=\exp\{ \half\Phi_i\} \left[ E^+ +\c D _+\Phi_i E^z\right],\quad
E^{-\al i}=\exp \{ \half\Phi_i\} \left[ E^- +\c D _-\Phi_i E^{\overline
 z}\right]\cr
E^{\al i+\al j}&=\exp \{ \half\left(\Phi_i+\Phi_j\right)\}  E^{z}\times
\cases{2 &$i\neq j$\cr 1&$i=j$\cr},\quad
E^{-\al i-\al j}=\exp \{ \half\left(\Phi_i+\Phi_j\right)\} E^{\overline z}
\times\cases{2 &$i\neq j$\cr 1&$i=j$\cr}\cr
E^\gamma& =0\ \ {\rm if}\ |{\rm height }(\gamma)|\geq3\cr}}
where $( E^A,\Omega)$ describes an arbitrary two dimensional
supergeometry.
We set the H-connections to
\eqn\ssetconn{i\sum _j \Omega ^j k_{ji}=\half \Omega \rho _i
+ E^A J_A{}^B \c D _ B\Phi_i}
where the supercomplex structure $J_A{}^B$ is defined as
\eqn\scomplex{J_A{}^B = \delta_A{}^B\times
\cases{+i&$A=z,+$\cr-i&$A=\overline z,-$\cr}}
The flatness condition $\c F _{\pm \pm}=0$ evaluated on \sansatz\
and \ssetconn\ reduces to the  torsion constraints of ordinary
supergravity and $\c F _{+-} =0$ yields Toda field equations in the
presence of N=1 supergravity.
\eqn\stodaeq{\c D _-\c D _+ \Phi_i+\sum _j \exp \{\Phi_j\}
k_{ji}-{i\over2} R_{+-}\rho _i=0}
The remaining flatness conditions on $\c F$ vanish by derivatives of
the torsion constraints and the Toda field equations.

The Lax pair is again easily identified as
the equation for parallel transport under the G--connection
\eqn\slaxpair{\left(\c D ^{(0)}_ A+\c A_A\right)\Psi=0}
or in terms of U(1)--frame index notation, we have the
equations for the components $A=\pm$
\eqn\sgaugefields{\eqalign{
(\c D_+ ^{(0)} + \sum _i \Omega _+ ^i h _i +
\exp \{ \half \Phi _i \} x_{\alpha _i}) \Psi& =0\cr
(\c D_- ^{(0)} + \sum _i \Omega _- ^i h _i
+ \exp \{ \half \Phi _i \} x_{-\alpha _i})\Psi&=0}}
The compatibility of the Lax pair \slaxpair\ for $A=+,-$
reduces to the Toda equations \stodaeq, using the definition of $\Omega ^i$
in \ssetconn.
Using the equation \slaxpair\ for $A=+$ twice, we find
$\c A_z$ and $\c A_{\overline z}$
\eqn\azeq{\c A_z=\c D ^{(\half)}_+ \c A_++\c A_+\wedge \c A_+,\qquad
\c A_{\overline z}=\c D ^{(-\half)} _- \c A_-+\c A_-\wedge \c A_-}
This, of course, agrees with the previous expression for $\c A$
in \sansatz.
Then using the Toda equation \stodaeq, the torsion constraints
of supergeometry and the Jacobi identity, we find that
the equations \slaxpair\ for all indices $A$ are compatible.

In  the case of supergroups for which all simple roots
cannot be chosen to be odd, we may only couple the Toda theory
to gravity.
Gravity is embedded in supergravity by setting the gravitino and auxiliary
fields to zero in the Wess--Zumino gauge
\RMP
\eqn\embed{\c D _+ ^{(n)}=\partial_\theta+\theta D_ z ^{(n)}
-{ i \over 2} \theta \bar \theta \omega _z \partial _{\bar \theta},\qquad
\c D _- ^{(n)} =\partial_{\overline\theta}+\overline\theta D_{\bar  z}
^{(n)}
-{i \over 2} \theta \bar \theta \omega _{\bar z} \partial _\theta }
where $D_a ^{(n)}$ was defined in \der.
The gauge field is of the following form
\def\oddroots{\Delta_s^{odd}}
\def\evenroots{\Delta_s^{even}}
\eqn\sconnpm{\eqalign{
\c A_+&=\sum _i \Omega _+ ^i h _i
 + \sum_{\al i\in\oddroots} \exp \{ \half \Phi _i \} x_{\alpha _i}
 + \sum_{\al i\in\evenroots}\theta \exp \{ \half \Phi _i \} x_{\alpha _i}\cr
\c A_-&=\sum _i \Omega _- ^i h _i
 + \sum_{\al i\in\oddroots} \exp \{ \half \Phi _i \} x_{-\alpha _i}
 +\sum_{\al i\in\evenroots} \overline\theta
  \exp\{ \half \Phi _i \} x_{-\alpha _i}\cr}}
where $\oddroots,\evenroots$ denote the set of positive odd and
even simple roots, respectively.
The H--connections $\Omega^i$ may be reexpressed in terms of $\Phi_i$
by exactly the same relation as before, namely equation \ssetconn.
The rest of the gauge field, $\c A_a$
 are defined using the relation \sgaugefields.

The condition $\c F_{+-}=0$ is equivalent to the Toda equation coupled
to gravity,
\eqn\sstodaeq{\c D_-\c D _+\Phi_i
+\sum_{\al j\in\oddroots} \exp \{\Phi_j\} k_{ji}
+\sum_{\al j\in\evenroots} \theta\overline\theta\exp \{ \Phi_j\}k_{ji}
-{i\over2}R_{+-} \rho _i =0}
The torsion constraints, the Toda field equations
and the Jacobi identity  guarantee that
the other components of $\c F_{AB}=0$ are satisfied.

The B\"acklund transformations may be obtained from the Lax equations
in analogy with the the methods used in the purely bosonic case.
Since one has to divide by the $\Psi$--field corresponding
to the highest weight, we must consider representations
in which the highest weight has even grading.
This is always possible, since the Lax equations are linear.
The relation between $\c A _a$ and $\c A _\alpha$,
\azeq, is automatically satisfied by the Lax equation,
\slaxpair, since $\c A = \c D \Psi \Psi ^{-1}$.

The action for the Toda field theory associated with a super Lie
algebra may be obtained when the Cartan
matrix is symmetric and invertible
\eqn\saction{S_\Phi=\int d^2\xi d^2\theta E\left [\sum _{i,j}
\{ \half l_{ij} D _+ \Phi _i D _{-} \Phi _j
-{i\over2}R_{+-} l_{ij} \rho _i \Phi _j\}
+\sum_{\al i\in\oddroots} \exp\{ \Phi _i\}
+\sum_{\al i\in\evenroots} \theta\overline\theta\exp\{ \Phi _i\}
\right ] }
This action is written in superfield language, but is (locally)
supersymmetric only when $\evenroots$ can be chosen to be empty.
\newsec{Summary}
In this paper, we have shown that Toda field theories coupled
to gravity or supergravity provide a class of solutions
to the W--gravity equations of motion.
Also, we obtained the Lax pair for the Toda system
on an arbitrary (super) Riemann surface.
This was shown for arbitrary Lie algebras and  superalgebras,
including the infinite dimensional ones.
It is natural to expect that Toda field theory can in fact be
coupled to an arbitrary W--gravity, providing  an integrable system as well.

Let us point out that the presence of (global) W--symmetries
\ref\zamo{A.B. Zamolodchikov, Theor. Math. Phys. 65 (1985) 1205 }
in Toda field systems was discussed in
\ref\bilal{
A. Bilal and J.-L. Gervais, Phys. Lett. B206 (1988) 412; Nucl. Phys.
B314 (1989) 646 \nextline
P. Mansfield and B. Spence, Nucl. Phys. B362 (1991) 294 }.
Also, some work on the relation between W--gravity and
Toda field theory has been presented in
\ref\hull{
C.M. Hull, Nucl. Phys. B364 (1991) 621; Phys. Lett. B269 (1991) 257
\nextline
C.N. Pope, ``Lectures on W-Algebras and W--gravity",
CTP--TAMU--103--91 (1991)\nextline
A.~ Gerasimov, A.~ Levin, A.~ Marshakov,  Nucl.~ Phys.~ B360 (1991) 537\nl
J.-M.~Lina, P.K.~Panigrashi, U.~Montr\'eal preprint, UdeM--LPN--TH--59 (1991)}
\listrefs
\vfil\eject\end